\documentstyle[11pt,appb1,epsfig,amssymb]{article}

\begin{document}
\begin{flushright}
CERN-TH/98-151 \\ 
\end{flushright}
\vspace*{0.1cm}
\begin{center}
{\large{$Q^2$-EVOLUTION OF SPIN-DEPENDENT \\ PARTON DENSITIES}} \\ 
\vspace*{0.4cm}
\renewcommand{\thefootnote}{\fnsymbol{footnote}}
\setcounter{footnote}{0}
{\sc Werner Vogelsang\footnote{Invited talk 
presented at the `Cracow Epiphany Conference on Spin Effects in
Particle Physics and Tempus Workshop', Jan. 9--11, 1998, Cracow, Poland}}
\renewcommand{\thefootnote}{\fnsymbol{roman}}
\setcounter{footnote}{0} \\
Theory Division, CERN, CH-1211 Geneva 23, Switzerland
\end{center}
\begin{abstract}
We discuss the NLO evolution of quark transversity densities and of the
parton distribution function for linearly polarized gluons in a linearly 
polarized hadron. A supersymmetric relation between the NLO evolution
kernels for transversity and for linear polarization is found. 
We also study the implications of NLO evolution for Soffer's inequality
and the prospects of measuring transversity densities in polarized
Drell--Yan at RHIC. 
\end{abstract}
\vspace*{-0.4cm}  
\section{Introduction}
Experimentally, the vast majority of data sensitive to parton densities
have been taken without fixing the polarization of the initial beams 
or the target. The densities extracted in this way are usually 
refered to as the `unpolarized' parton distributions $f (x,Q^2)$ 
($f=q,\bar{q},g$).

Within roughly the last decade, also more and more data have become available 
that are sensitive to the `longitudinally' polarized (`helicity--weighted')
parton densities of the nucleon. The tool to obtain such information
has (almost exclusively) been deep--inelastic scattering (DIS) of
longitudinally polarized leptons and nucleons. The spin--asymmetry
measured in such reactions is related to the probability for finding a 
certain parton--type with positive helicity in a nucleon of positive 
helicity {\em minus} the probability for finding it with negative helicity. 
These densities, denoted as $\Delta f (x,Q^2)$ ($f=q,\bar{q},g$), contain 
information different from that contained in the more familiar unpolarized 
ones.

For a {\em transversely} polarized spin--$1/2$ hadron one can define a 
further quark density~\cite{rs,ji,am} in very much the same way as the 
longitudinally 
polarized quark distributions, by taking differences of probabilities for 
finding quarks with transverse spin aligned and anti--aligned with the 
transverse hadron spin. These densities are called `transversity'
densities and are denoted by $\Delta_T f(x,Q^2)$. It turns
out that -- unlike the situation for unpolarized and longitudinally 
polarized densities -- there is {\em no} gluonic analogue of quark 
transversity~\cite{ji}. This is due to angular momentum conservation: 
transversity densities are related to helicity--flip amplitudes. A gluonic 
helicity--flip amplitude would require the hadron to absorb two units 
of helicity, which no spin--$1/2$ target can do.

The transversity distributions are completely unmeasured
so far since they cannot be directly accessed in DIS. It seems certain, 
however, that measurements of the $\Delta_T f$ will be attempted at 
the future polarized proton-proton collider RHIC at Brookhaven~\cite{rhic}. 
The most suitable candidate for such measurements is believed to be
Drell--Yan dimuon production~\cite{rs}.

The $\Delta_T f$ complete the twist--2 sector of parton densities 
of spin--$1/2$ hadrons. Nevertheless, we have not
yet depleted the full set of parton densities that can be defined if spin
is taken into account. There finally is a further spin--dependent {\em gluon} 
distribution that to some extent can be regarded as the gluonic counterpart 
of quark transversity. Unlike the `helicity' density $\Delta g$
that describes {\em circular} polarization of the gluon, it is 
encountered if the gluon is {\em linearly} polarized~\cite{am,jaffe,deld}.
The density is denoted by $\Delta_L g(x,Q^2)$ and exists only in a 
linearly polarized hadron (or photon~\cite{deld}),
which therefore has to have spin $\geq 1$. There is no quark distribution 
in this case~\cite{jaffe}. Even though a measurement of $\Delta_L g$ does 
not seem very realistic at the moment, it possesses some interesting 
theoretical aspects which justify its analysis.

Table~1 summarizes the parton densities we have defined. 
\begin{table}[b]
\renewcommand{\arraystretch}{1.6}
\begin{center}
\begin{tabular}{|c|c|c|} \hline \hline
polarization & quarks & gluons \\ \hline \hline
unpolarized & $q\equiv q_+^+ + q_+^-\equiv q_{\uparrow}^{\uparrow} + 
q_{\uparrow}^{\downarrow}$ & $g\equiv g_+^+ + g_+^-
\equiv g_x^x + g_y^x$ \\ \hline
long. polarized & $\Delta q = q_+^+ - q_+^-$ & $\Delta g = g_+^+ - g_+^-$ 
\\ \hline
transversity & $\Delta_T q = q_{\uparrow}^{\uparrow} - 
q_{\uparrow}^{\downarrow}$ & {\bf ---} \\ \hline
linearly pol. glue & {\bf ---} & $\Delta_L g = g_{\hat{x}}^{\hat{x}} - 
g_{\hat{y}}^{\hat{x}}$ \\ \hline
\hline
\end{tabular}
\vspace*{1cm}
\caption{\sf List of twist--2 quark and gluon densities including 
spin--dependence. We have suppressed the ubiquitous argument $(x,Q^2)$ of 
the densities. Note that `$q$' always runs over quarks as well as over 
antiquarks. Labels $+,-$ denote helicities, $\uparrow,\downarrow$ 
transverse polarizations, and $\hat{x}$ ($\hat{y}$) stands for linear 
polarization along the $x$ ($y$) axis, where the particle is moving along 
the $z$--direction. Subscripts refer to partons and superscripts to 
the parent hadron.}
\vspace*{-0.5cm}
\end{center}
\end{table}

It is important to realize that each set of parton densities in Tab.~1
(i.e., each of the rows of Tab.~1) is subject to its own set of evolution 
equations. For instance, the evolution of the longitudinally polarized 
densities proceeds independently from that of the unpolarized partons,
and so forth. In this way, we are led to introducing separate sets of 
evolution kernels (splitting functions) for each type of polarization. 

The $Q^2$--evolution of the unpolarized densities has been worked out up to
NLO accuracy of QCD already a long time ago~\cite{ap,splitunp,cfp}, and it has
become standard since about ten years to analyse the unpolarized data 
within the NLO framework. The LO evolution of the $\Delta f (x,Q^2)$,
the $\Delta_T f (x,Q^2)$, and of $\Delta_L g (x,Q^2)$ has also been known
for a long time~\cite{ap,ahro,am}, whereas the derivation of the NLO evolution
kernels for longitudinally polarized partons has been a more recent
development~\cite{mvn,wv}. Very recently, the NLO splitting functions
for transversity were derived within three independent
calculations~\cite{wvt,j1}. In this paper, we will discuss the
NLO evolution of the $\Delta_T f$, and its implications for an inequality
between the $f$, $\Delta f$ and $\Delta_T f$ derived by Soffer~\cite{soffer}. 
We will also for the first time present the NLO evolution 
kernel for $\Delta_L g$.
\section{NLO evolution equations}
Since there are no gluons involved in the case of the transversity
distributions, their evolution equations reduce to simple non-singlet type
equations. Introducing 
\begin{equation} \label{pm}
\Delta_T q_{\pm}^n (Q^2) \equiv \Delta_T q^n (Q^2) \pm \Delta_T 
\bar{q}^n (Q^2) \; ,
\end{equation}
where the Mellin moments are defined by
\begin{equation} 
\Delta_T q^n (Q^2) \equiv \int_0^1 x^{n-1} \Delta_T q^n (x,Q^2) \; ,
\end{equation}
one has the evolution equations (see, for example, \cite{ev})
\begin{equation} \label{evol3}
\frac{d}{d\ln Q^2} \Delta_T q_{\pm}^n (Q^2) = \Delta_T P_{qq,\pm}^n
(\alpha_s (Q^2)) \Delta_T q_{\pm}^n (Q^2)
\end{equation}
for all flavours. The splitting functions $\Delta_T P^n_{qq,\pm}
(\alpha_s (Q^2))$ are taken to have the following
perturbative expansion:
\begin{equation} \label{expan}
\Delta_T P_{qq,\pm}^n (\alpha_s) =
\left( \frac{\alpha_s}{2\pi} \right) \Delta_T
P_{qq}^{(0),n} + \left( \frac{\alpha_s}{2\pi} \right)^2 \Delta_T
P_{qq,\pm}^{(1),n} + \ldots  \; .
\end{equation}
As indicated, $\Delta_T P_{qq,+}$ and $\Delta_T P_{qq,-}$ are equal at LO.

Similarly to~(\ref{evol3}) one has for linearly polarized gluons:
\begin{equation} \label{evol4}
\frac{d}{d\ln Q^2} \Delta_L g^n (Q^2) = \Delta_L P_{gg}^n
(\alpha_s (Q^2)) \Delta_L g^n (Q^2) \; ,
\end{equation}
where $\Delta_L P_{gg}^n$ has a peturbative expansion analogous 
to~(\ref{expan}).
These NLO evolution equations can be easily solved; for details see, 
for instance, \cite{wvt}.
\section{Calculation of splitting functions}
Historically, the evolution kernels for parton densities have been 
calculated within two rather different methods. On the one hand, the 
very powerful, yet quite formal, technique of the 
`Operator--Product--Expansion' (OPE) has been applied. Here the evolution 
kernels are derived as anomalous dimensions of matrix elements of 
local operators. The other method is more intuitive and relies on 
parton model ideas and on the factorization of mass singularities in a 
physical (ghost--free) gauge~\cite{egmpr,cfp}. This is the method we will use.
Very roughly, the strategy goes as follows: The NLO splitting functions
are related to the residues of the collinear singularities in the
{\em next--to}--NLO (NNLO) calculation of a partonic subprocess
cross section for a `physical' process. Being the pole part of this cross
section, the singular terms  are manifestly gauge--{\em in}dependent. 
This means that we can employ a gauge in their calculation that simplifies 
the extraction of the mass singularities as much as possible. As was 
shown in~\cite{egmpr}, working in a physical gauge, like the light--cone
gauge, reduces the number of singular graphs at a given order significantly.
Thus, a full NNLO calculation is clearly not required. In particular, only 
ladder--like diagrams, corresponding to a parton cascade, contribute 
to the pole part in such gauges, so that the cross section $\hat{\sigma}$ 
for any physical partonic process can be diagrammatically expanded into 
a sum of ladders of `two--particle--irreducible' kernels that are 
individually finite. Mass singularities only occur on the lines 
connecting the kernels (i.e., on the `sides' of the ladders) and can 
therefore be projected out easily using a projector ${\cal P}$ that 
converts these lines into on--shell physical states.
Thus, the factorization of mass singularities can be achieved: the cross
section $\hat{\sigma}$ decomposes into a finite part $\hat{\sigma}_F$,
and a function $\Gamma$ that contains all the collinear singularities and 
is universal since no process--dependence is left over in it. 
The function $\Gamma$ is then directly related to the splitting functions 
one is looking for~\cite{egmpr,cfp}.

\newcommand{\slsh}{\rlap{$\;\!\!\not$}}
For the case of transversity, the projector $\Delta_T {\cal P}$ is found to be
\begin{equation} \label{g52}
\Delta_T {\cal P} \sim \frac{1}{4 n \cdot k} \slsh{n} \slsh{s} \gamma_5 \; ,
\end{equation}
where $k$ is the momentum of the particle emerging from the top of a kernel,
$n$ is the gauge vector with $n\cdot A=0$, $n^2=0$ for the light--cone
gauge, and $s$ is the transverse spin vector. More details on the calculation 
in the transversity case can be found  in~\cite{wvt}.

For linearly polarized gluons one has
\begin{equation} \label{proj}
\Delta_L {\cal P}_{\hat{x}\hat{y}}^{\alpha\beta} \equiv \frac{1}{2} \left(
\epsilon_{\hat{x}}^{\alpha} \epsilon_{\hat{x}}^{\beta} - 
\epsilon_{\hat{y}}^{\alpha} \epsilon_{\hat{y}}^{\beta}  \right) \; ,
\end{equation}
where the polarization vectors
\begin{equation}
\epsilon_{\hat{x}} \equiv (0,1,0,\ldots,0) \; , \;\;\; 
\epsilon_{\hat{y}} \equiv (0,0,1,0,\ldots,0) 
\end{equation}
have non--vanishing entries only in the $x$ and $y$ components, respectively.
A more detailed account of our calculation of $\Delta_L P_{gg}^{(1)}$
will be given in a future publication~\cite{fut}.
\section{Results}
Our results for $\Delta_T P_{qq,\pm}^{(1)} (x)$ can be found in~\cite{wvt}
and need not be repeated here. The one for $\Delta_L P_{gg}^{(1)}$ is
new and reads in the $\overline{{\mbox{\rm MS}}}$ scheme
\begin{eqnarray} \label{pgglnlo}
\Delta_L P_{gg}^{(1)} (x) &=& N_C^2 \Bigg[ \left( \frac{67}{18} + \frac{1}{2}
\ln^2 x- 2 \ln x \ln (1-x) - \frac{\pi^2}{6} \right) \delta_L P_{gg}^{(0)}(x)
\nonumber \\
&&\hspace*{1cm}+\frac{1-x^3}{6 x} + 
S_2 (x) \delta_L P_{gg}^{(0)}(-x) + \left( 
\frac{8}{3}+3 \zeta (3) \right) \delta (1-x) \Bigg] \nonumber \\
&+&N_C T_f \left[ -\frac{10}{9} \delta_L P_{gg}^{(0)}(x) + 
\frac{1 - x^3}{3x} -\frac{4}{3} \delta (1-x) \right]  \nonumber \\ 
&-&C_F T_f \left[ 2 \frac{1 - x^3}{3x} + \delta (1-x) \right] \; ,
\end{eqnarray}
where
\begin{eqnarray} 
\delta_L P_{gg}^{(0)} (x) &=& \frac{2x}{(1-x)_+} \; , \\
S_2(x) &=& \int_{\frac{x}{1+x}}^{\frac{1}{1+x}} \frac{dz}{z} 
\ln \big(\frac{1-z}{z}\big) \nonumber \\
&=& -2 {\rm Li}_2 (-x)-2 \ln x \ln (1+x)+\frac{1}{2} \ln^2 x-
\frac{\pi^2}{6} \; ,
\end{eqnarray}
with ${\rm Li}_2 (x)$ being the dilogarithm. Furthermore, we have
as usual $C_F=4/3$, $N_C=3$, and $T_f=n_f/2$, $n_f$ being the
number of active flavours. Finally, $\zeta (3) \approx 1.202057$.
We note that in
\cite{deld} a result for the part $\sim C_F T_f$ in (\ref{pgglnlo}) at $x<1$ 
was presented, which corresponds to the two--loop splitting function for
transitions between linearly polarized gluons and {\em photons}. 
The result of \cite{deld} disagrees with ours by a an additional 
factor $1/x$ in \cite{deld}. The calculation of \cite{deld} therefore has to 
be in error, since a behaviour $\sim 1/x^2$ of the splitting function --
more singular than the unpolarized one -- cannot be correct.

There are two aspects of the result (\ref{pgglnlo}) that deserve
further attention. Firstly, the small--$x$ behaviour of the splitting
function for linear polarization changes quite dramatically when going from
LO to NLO. At LO, the splitting function behaves $\sim x$ as
$x\rightarrow 0$. At NLO, there are terms $\sim 1/x$ in the splitting
function (as one also encounters in the unpolarized case); we have
\begin{equation} \label{smallx1}
\Delta_L P_{gg}^{(1)} (x) \approx \frac{1}{3x} \left[ 
\frac{1}{2} N_C^2 + N_C T_f - 2 C_F T_f \right] + {\cal O} (x) 
\;\;\;\;\; (x\rightarrow 0) \; .
\end{equation}
We note that all logarithmic terms $\sim x \ln^2 x$ cancel out in this 
limit. 

The other interesting point concerns the `supersymmetric' limit 
$C_F=N_C=2 T_f\equiv N$ \cite{bukh} which has already been investigated 
for the unpolarized and the longitudinally polarized NLO splitting functions
in~\cite{ant,mvn,wv} and for the `time--like' ones in~\cite{svtl}.
One first of all finds that in the supersymmetric limit the LO
splitting functions for quark transversity and for linearly polarized
gluons become equal~\cite{am,deld}:
\begin{equation}
\Delta_T P_{qq}^{(0)} (x) = N \left[ \frac{2x}{(1-x)_+} + \frac{3}{2}
\delta (1-x) \right] = \Delta_L P_{gg}^{(0)} (x) \; .
\end{equation}
Thus, we have found the supersymmetric `counterpart' of transversity:
linear polarization (see also~\cite{bukh}). This nicely completes the 
supersymmetry relations found in the case of unpolarized and
longitudinally polarized parton densities and fragmentation functions.
The fact that two rather different polarizations are linked here is
not too surprising as both of them are transverse in a certain sense.

To see whether the supersymmetric relation also holds at NLO, we
have to transform the splitting functions to a regularization
scheme that respects supersymmetry, dimensional reduction. The procedure
here follows closely that discussed in~\cite{wv,svtl}. The task is
simplified by the fact that there are no parts $\sim \epsilon$ in the 
$d$--dimensional LO splitting functions for transversity or linearly
polarized gluons at $x<1$. Such terms -- which are always absent in
dimensional reduction, but can be present for dimensional 
{\em regularization} -- are the reason why, for instance, the longitudinally
polarized NLO splitting functions do not directly satisfy the supersymmetric 
relation in dimensional regularization, but have to be 
transformed to dimensional reduction first (see \cite{wv} for a more 
detailed discussion). The fact that the LO $\epsilon$--terms are absent at 
$x<1$ in our case, means that our final results for the NLO transversity 
splitting function $\Delta_T P_{qq,+}^{(1)}$ of~\cite{wvt}
and for the NLO splitting function for linearly polarized gluons,
Eq.~ (\ref{pgglnlo}), already coincide for $x<1$ with their respective
$\overline{{\mbox{\rm MS}}}$ expressions in dimensional {\em reduction}. 
We can therefore immediately compare these expressions in the supersymmetric 
limit. Indeed, we find for $C_F=N_C=2 T_f\equiv N$:
\begin{equation}
\Delta_T P_{qq,+}^{(1)} (x) \equiv \Delta_L P_{gg}^{(1)} (x) \;\;\;\;
(x<1) \; .
\end{equation}
The satisfaction of the supersymmetric relation at $x=1$ is trivial;
see~\cite{wv}, where the appropriate factorization scheme transformation
to dimensional reduction at $x=1$ is given.

We thus find that indeed the supersymmetric relation between 
the trans\-vers\-ity splitting function and the evolution kernel for 
linearly polarized gluons is also verified at NLO. We note that
the relation involves as expected the `$+$-combination' 
$\Delta_T P_{qq,+}^{(1)}\equiv \Delta_T P_{qq}^{(1)} +
\Delta_T P_{q\bar{q}}^{(1)}$ which corresponds to `singlet' evolution.
Finally, also note the miraculous cancellation of the terms $\sim 1/x$ in 
$\Delta_L P_{gg}^{(1)}$ in the supersymmetric limit (see Eq.~(\ref{smallx1}))
-- after all, such terms are not present in $\Delta_T P_{qq,+}^{(1)}$. 
\section{Soffer's inequality and the transversely polarized Drell--Yan
process}
The unpolarized, longitudinally and transversely polarized quark
distributions ($q$, $\Delta q$, $\Delta_T q$) of
the nucleon are expected to obey the rather interesting relation
\begin{equation} \label{soffer1}
2 |\Delta_T q(x)| \leq q(x) + \Delta q(x) \label{soffer}
\end{equation}
derived by Soffer \cite{soffer}. 
It has recently been clarified that Soffer's inequation is 
preserved by leading order (LO) QCD evolution, i.e.\ if (\ref{soffer}) 
is valid at some scale $Q_0$, it will also be valid at $Q>Q_0$ \cite{baro}.
To NLO the situation is not as simple. The parton distributions are now 
subject to the choice of the factorization scheme which one may fix
independently for $q$, $\Delta q$ and $\Delta_T q$. One can therefore always 
find `sufficiently incompatible' schemes in which a violation of
(\ref{soffer}) occurs. However, it was shown in \cite{wvt}
with analytical methods that the inequality for valence 
densities is preserved by NLO QCD evolution in a certain 
`Drell-Yan scheme' in which the NLO cross sections for dimuon production
maintain their LO forms, and also in the $\overline{\mbox{MS}}$ 
scheme. In this section, which is taken from~\cite{om}, we shall show
numerically that Soffer's inequation for sea quarks is also preserved
under NLO ($\overline{\mbox{MS}}$) evolution. 
We note that there is also a recent analytical proof
of the preservation of Soffer's inequality~\cite{soff1} at NLO.

For our study, we will assume saturation of Soffer's inequality at the
input scale for parton evolution. Our choice for the
r.h.s. of (\ref{soffer1}) will then be the NLO $\overline{\mbox{MS}}$ 
radiative parton model inputs for $q(x,Q_0^2)$ of \cite{glre2} and for the 
longitudinally polarized $\Delta q(x,Q_0^2)$ of the `standard' scenario 
of \cite{glre3} at $Q_0^2=\mu_{NLO}^2=0.34$
GeV$^2$. For simplicity we will slightly deviate from the actual $q(x,Q_0^2)$
of \cite{glre2} in so far as we will neglect the breaking of SU(2) in the 
input sea quark distributions originally present in this set. This seems
reasonable as SU(2) symmetry was also assumed for the $\Delta \bar{q}
(x,Q_0^2)$ of \cite{glre3}, which in that case was due to the fact that
in the longitudinally polarized case there are no data yet that could
discriminate between $\Delta \bar{u}$ and $\Delta \bar{d}$. We therefore
prefer to assume $\Delta_T \bar{u} (x,Q_0^2)=\Delta_T \bar{d} (x,Q_0^2)$ also
for the transversity input.

\begin{figure}[ht]
\epsfysize9cm
\hspace*{-0.2cm}
\leavevmode\epsffile{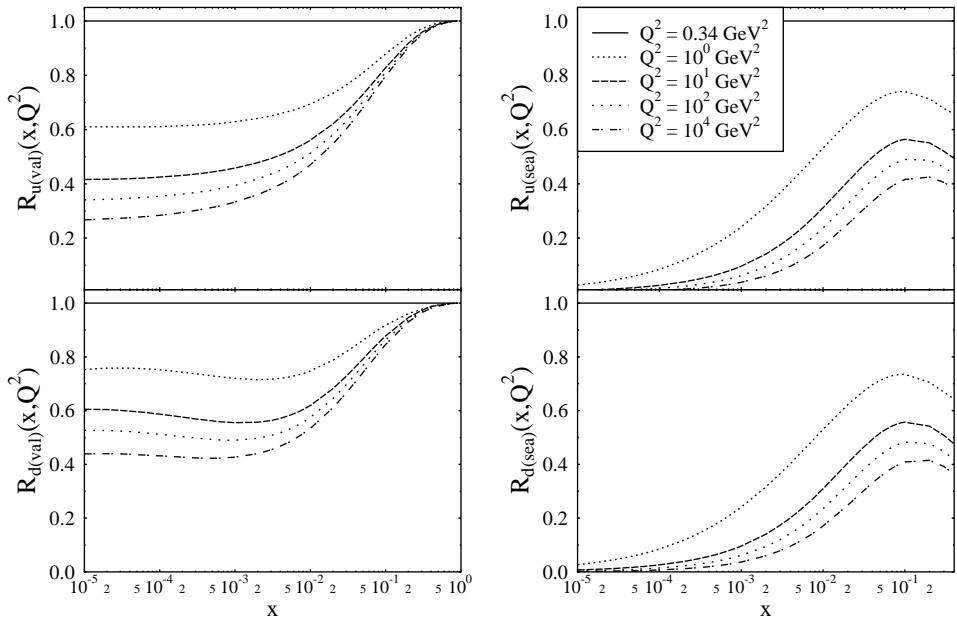}
\caption{\label{fig1}The ratio $R_q(x,Q^2)$ as defined in (\ref{req}) 
for $q=u_v,\bar{u},d_v,\bar{d}$ and several fixed values of $Q^2$.}
\end{figure}

In order to numerically check the preservation of (\ref{soffer}), 
Fig.~\ref{fig1} shows the ratio
\begin{equation}
R_q(x,Q^2) = \frac{2|\Delta_T q(x,Q^2)|}{q(x,Q^2)+\Delta q(x,Q^2)}
\label{req}
\end{equation}
as a function of $x$ for several different $Q^2$ values for $q=u_v=u_-$, 
$\bar{u}=(u_+-u_-)/2$, $d_v=d_-$, $\bar{d}=(d_+-d_-)/2$ (cf. Eq.~(\ref{pm})).
If NLO evolution preserves Soffer's inequality, then $R_q(x,Q^2)$ should not 
become larger than 1 for any $Q^2\geq Q_0^2$. As we already know from 
\cite{wvt} this is the case for the two valence 
distributions. Figure~\ref{fig1} confirms that Soffer's inequality is 
indeed also preserved for sea distributions. Furthermore, we see
in Fig.~\ref{fig1} that evolution leads to a strong suppression of 
$R_q(x,Q^2)$ at small values of $x$, in particular for the sea quarks. 

As is obvious from (\ref{soffer}), Soffer's inequality  only restricts the 
absolute value of the transversity distribution. Therefore, we are free to 
choose the sign when saturating (\ref{soffer}), and we have 
to check the results for the two distinct possibilities 
$\Delta_T q_v(x,Q_0^2)>0$,~$\Delta_T \bar{q}(x,Q_0^2)>0$ 
and $\Delta_T q_v(x,Q_0^2)>0$,~$\Delta_T \bar{q}(x,Q_0^2)<0$. 
Our results do not noticeably depend on the actual choice. Neither does
it matter whether we decide to saturate Soffer's inequality for
the {\em valence} densities $q_v$ or for the full quark 
($q\equiv q_v + \bar{q}$) distributions at the input scale.

One can utilise a saturated Soffer inequality to derive upper bounds
on the transverse double--spin asymmetry $A_{TT}$ to be measured in
transversely polarized Drell-Yan muon pair production at RHIC.
$A_{TT}$ is defined as $A_{TT}=d\delta \sigma/d\sigma$ where the
polarized and unpolarized hadronic cross sections are
\begin{equation}
d\delta\sigma \equiv \frac{1}{2} \left( 
d\sigma^{\uparrow\uparrow}-d\sigma^{\uparrow\downarrow}\right)\quad,
\quad\quad
d\sigma \equiv \frac{1}{2} \left( 
d\sigma^{\uparrow\uparrow}+d\sigma^{\uparrow\downarrow}\right)\quad.
\end{equation}
We employ the same input distributions as before, along with the same 
value for the initial scale $Q_0$. We choose $Q_F=Q_R=M$ for the 
factorization and renormalization scales, where $M$ is the 
invariant mass of the muon pair. Further details of the
calculation of the Drell--Yan cross section to LO and NLO can be
found in~\cite{om}.
\begin{figure}[ht]
\hspace*{0.2cm}
\epsfysize8cm
\leavevmode\epsffile{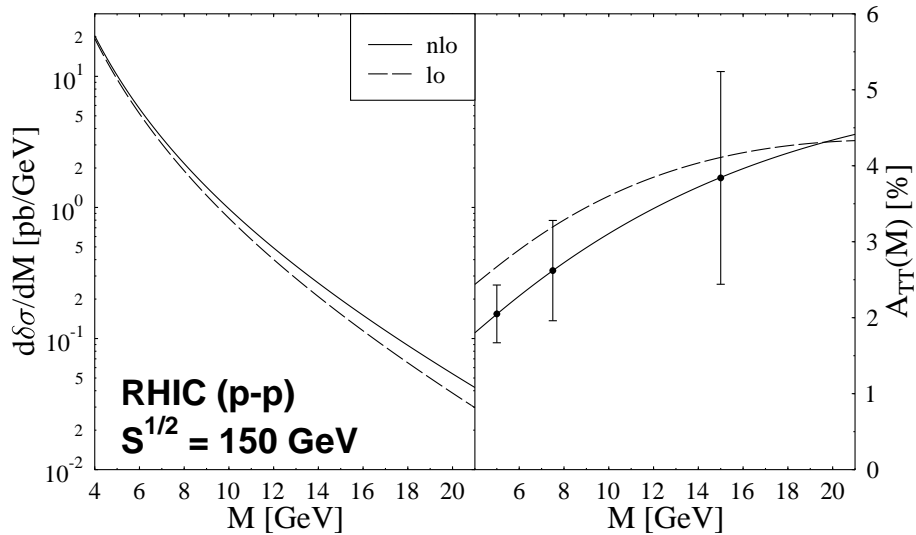}
\caption{\label{fig2} NLO and LO maximal polarized Drell-Yan cross sections
and asymmetries for RHIC at $\sqrt{S}=150\,\mbox{GeV}$.
The error bars have been calculated assuming ${\cal L}=240\,\mbox{pb}^{-1}$,
70\% beam polarization and 100\% detection efficiency.}
\end{figure}

Figure~\ref{fig2} shows the transversely polarized $pp$ cross section and 
the `maximal' double--spin asymmetry $A_{TT}$ at LO and NLO for
$\sqrt{S}=150\;\mbox{GeV}$, corresponding to the RHIC $pp$
collider. We note that the region $9$ GeV $\lesssim M \lesssim 11$ GeV
will presumably not be accessible experimentally since it will be 
dominated by muon pairs from bottomonium decays.
The predicted maximal asymmetry is of the order of a few
per cent. The expected statistical error bars have been included in the
plot; they have been calculated for $70\%$ beam polarization, an
integrated luminosity ${\cal L}=240\;\mbox{pb}^{-1}$, and $100\%$
detection efficiency.
One concludes that asymmetries of this size should be well measurable
at RHIC.  Similar asymmetries are found~\cite{om} for another conceivable
experimental situation, namely a proposed fixed--target experiment at HERA
that would utilise the possibly forthcoming polarized 820 GeV HERA
proton beam on a transversely polarized target,
resulting in $\sqrt{S}=40\;\mbox{GeV}$.

\begin{figure}[ht]
\hspace*{0.2cm}
\epsfysize8cm
\leavevmode\epsffile{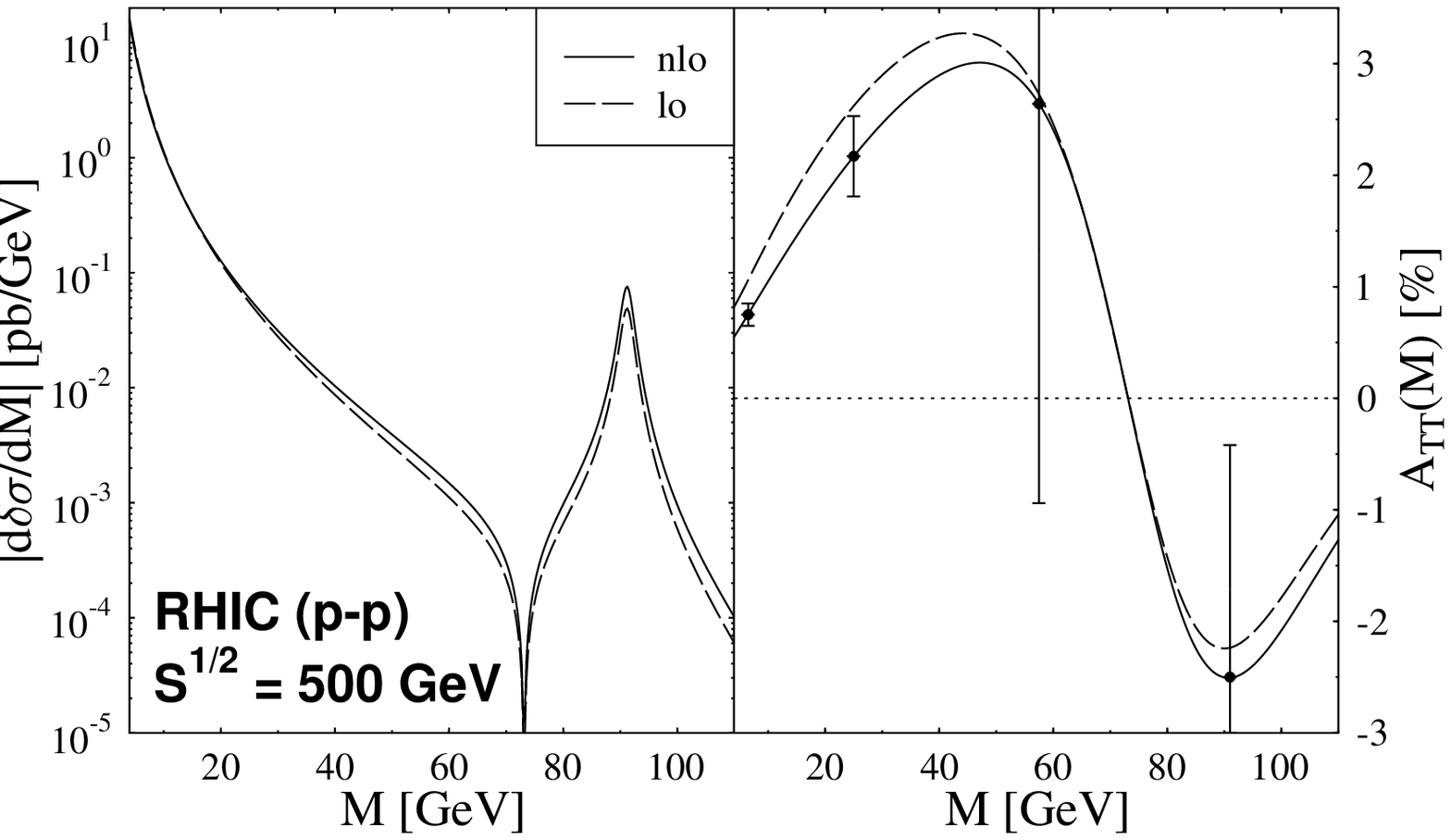}
\caption{\label{fig3}As in Fig.~\ref{fig2}, but for $\sqrt{S}=500\,\mbox{GeV}$
and ${\cal L}=800\,\mbox{pb}^{-1}$.}
\end{figure}

Figure~\ref{fig3} shows similar results for the high-energy end of RHIC,
$\sqrt{S}=500\;\mbox{GeV}$, where the integrated luminosity is expected 
to be ${\cal L}=800\;\mbox{pb}^{-1}$. It turns out that the asymmetries
become smaller as compared to the lower energies, but thanks to the
higher luminosity the error bars become relatively smaller as well, at least 
in the region $5\;\mbox{GeV}\lesssim M \lesssim 25\;\mbox{GeV}$ where the 
errors are approximately 1/10 of the maximal asymmetry. One can also clearly 
see in Fig.~\ref{fig3} the effect of $Z$ exchange and the $Z$ resonance.

A comparison of the LO and NLO results in Figs.~\ref{fig2} and \ref{fig3} 
answers one key question concerning the transversely polarized Drell-Yan 
process: our predictions for the maximal $A_{TT}$ show good perturbative 
stability, i.e.\ the NLO corrections to the 
cross sections and $A_{TT}$ are of moderate size, albeit not negligible. 

\begin{figure}[ht]
\vspace*{-0.1cm}
\hspace*{0.2cm}
\epsfysize8cm
\leavevmode\epsffile{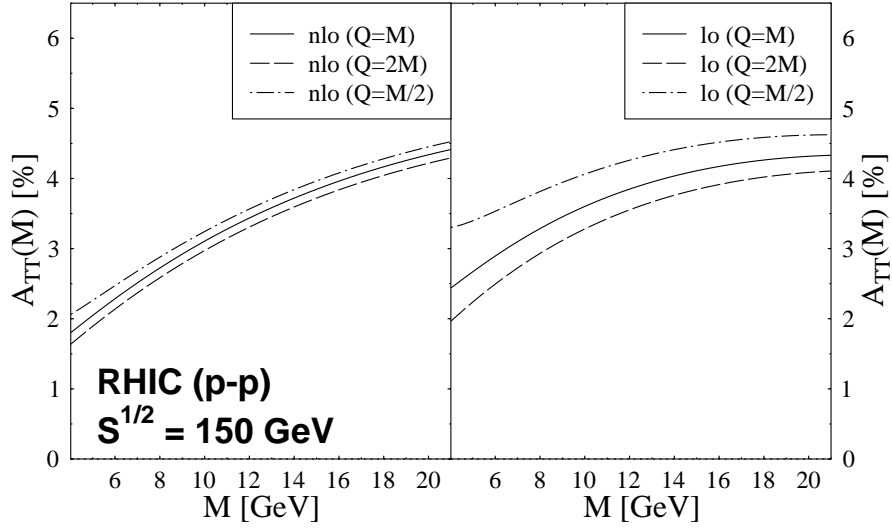}
\caption{\label{fig4}Scale dependence of the LO and NLO asymmetries at
$\sqrt{S}=150\,\mbox{GeV}$. The renormalization and factorization scales 
were chosen to be $Q_R=Q_F=M/2,M,2M$. The solid line is as in Fig.~\ref{fig2}.}
\end{figure}

Finally, we also find a significantly reduced scale dependence of the
results when going from LO to NLO. This is shown in Fig.~\ref{fig4} for 
the case $\sqrt{S}=150\;\mbox{GeV}$. We plot here the maximal asymmetry 
in LO and NLO, varying the renormalization and factorization scales in the  
range $M/2 \leq Q_F=Q_R \leq 2M$. One can see that already the LO 
asymmetry is fairly stable with respect to scale changes, which is in 
accordance with the finding of generally moderate NLO corrections. The NLO 
asymmetry even shows a significant improvement, so that $A_{TT}$ becomes
largely insensitive to the choice of scale. 

\noindent
{\bf Acknowledgement:} I am grateful to O.\ Martin, A.\ Sch\"{a}fer and 
M.\ Stratmann for a fruitful collaboration.
\newpage

%

\begin{thebibliography}{99}
\bibitem{rs} J.P.\ Ralston and D.E.\ Soper, Nucl. Phys. {\bf B152}, 109
(1979); R.L.\ Jaffe and  X.\ Ji, Phys. Rev. Lett. {\bf 67}, 
552 (1991); Nucl. Phys. {\bf B375}, 527 (1992); 
J.L.\ Cortes, B.\ Pire, and J.P.\ Ralston,
Z. Phys. {\bf C55}, 409 (1992).
\bibitem{ji} X.\ Ji, Phys. Lett. {\bf B289}, 137 (1992).
\bibitem{am} X.\ Artru and M.\ Mekhfi, Z. Phys. {\bf C45}, 669 (1990).
\bibitem{rhic} RHIC Spin Collab., D.\ Hill et al., Letter of intent 
RHIC-SPIN-LOI-1991, updated 1993; 
G.\ Bunce et al., Particle World {\bf 3}, 1 (1992).
\bibitem{jaffe} R.L.\ Jaffe and A.\ Manohar, Phys. Lett. 
{\bf B223}, 218 (1989).
\bibitem{deld} F.\ Delduc, M.\ Gourdin, E.G.\ Oudrhiri--Safiani,
Nucl. Phys. {\bf B174}, 157 (1980).
\bibitem{ap} G. Altarelli and G. Parisi, Nucl. Phys. {\bf B126},
298 (1977).
\bibitem{splitunp} E.G. Floratos, D.A. Ross, and C.T. Sachrajda,
Nucl. Phys. {\bf B129}, 66 (1977); E: {\bf 139}, 545 (1978);
Nucl. Phys. {\bf B152}, 493 (1979).
\bibitem{cfp} G. Curci, W. Furmanski, and R. Petronzio, Nucl. Phys.
{\bf B175}, 27 (1980); W. Furmanski and R. Petronzio, Phys. Lett.
{\bf 97B}, 437 (1980).
\bibitem{ahro} M.A.\ Ahmed and G.G.\ Ross, Nucl. Phys. {\bf B111}, 441 (1976).
\bibitem{mvn} R.\ Mertig and W.L.\ van Neerven, Z. Phys. {\bf{C70}}, 
637 (1996).
\bibitem{wv} W.\ Vogelsang, Phys. Rev. {\bf{D54}}, 2023 (1996);
Nucl. Phys. {\bf{B475}}, 47 (1996).
\bibitem{wvt} W.\ Vogelsang, Phys. Rev. {\bf{D57}}, 1886 (1998).
\bibitem{j1} S.\ Kumano and M.\ Miyama, Phys. Rev. {\bf D56}, 2504 (1997); \\
A.\ Hayashigaki, Y.\ Kanazawa and Y.\ Koike, 
Phys. Rev. {\bf D56}, 7350 (1997).  
\bibitem{soffer} J.\ Soffer, Phys. Rev. Lett. {\bf 74}, 1292 (1995).
\bibitem{ev} R.K.\ Ellis and W.\ Vogelsang, {\tt hep-ph/9602356} (unpublished).
\bibitem{egmpr} R.K.\ Ellis, H.\ Georgi, M.\ Machacek, H.D.\ Politzer, 
and G.G.\ Ross, Phys. Lett. {\bf 78B}, 281 (1978); Nucl. Phys.
{\bf B152}, 285 (1979).
\bibitem{fut} M.\ Stratmann and W.\ Vogelsang, in preparation.
\bibitem{bukh} A.P.\ Bukhvostov, G.V.\ Frolov, E.A.\ Kuraev, and L.N.\ 
Lipatov, Nucl. Phys. {\bf B258}, 601 (1985);
Yu.L.\ Dokshitser, Sov. Phys. JETP {\bf 46}, 641 (1987).
\bibitem{ant} I. Antoniadis and E.G. Floratos, Nucl. Phys. 
{\bf B191}, 217 (1981).
\bibitem{svtl} M.\ Stratmann and W.\ Vogelsang, Nucl. Phys. {\bf B496}, 41
(1997).
\bibitem{baro} V.\ Barone, Phys. Lett. {\bf B409}, 499 (1997).
\bibitem{om} O.\ Martin, A.\ Sch\"{a}fer, M.\ Stratmann and W.\ Vogelsang, 
Phys. Rev. {\bf D57}, 3084 (1998).
\bibitem{soff1} C.\ Bourrely, J.\ Soffer and O.V.\ Teryaev, 
                Phys. Lett. {\bf B420}, 375 (1998).
\bibitem{glre2} M.\ Gl\"uck, E.\ Reya and A.\ Vogt, Z. Phys. {\bf C67},
               433 (1995).
\bibitem{glre3} M.\ Gl\"uck, E.\ Reya, M.\ Stratmann,
                and W.\ Vogelsang, Phys. Rev. {\bf D53}, 4775 (1996).
\end{thebibliography}
\end{document}